\documentclass[conference]{IEEEtran}
\IEEEoverridecommandlockouts
\usepackage{cite}
\usepackage{amsmath,amssymb,amsfonts}
\usepackage{algorithmic}
\usepackage{graphicx}
\usepackage{textcomp}
\usepackage{cite}
\usepackage{verbatim}
\usepackage{hyperref}
\usepackage[caption=false,font=normalsize,labelfont=sf,textfont=sf]{subfig}
\usepackage{xcolor}
\usepackage{array}
\usepackage[ruled,linesnumbered]{algorithm2e}
\def\BibTeX{{\rm B\kern-.05em{\sc i\kern-.025em b}\kern-.08em
    T\kern-.1667em\lower.7ex\hbox{E}\kern-.125emX}}
\begin{document}

\title{Multi-Objective-Optimization Assisted Data Collection Framework for IoUT Based on Offline Reinforcement\\
}

\author{
\IEEEauthorblockN{
Yimian Ding\thanks{$^+$ These authors contribute equally to this work.}\IEEEauthorrefmark{1}$^{,+}$,
Xinqi Wang\IEEEauthorrefmark{2}$^{,+}$,
Jingzehua Xu\IEEEauthorrefmark{1},
Guanwen Xie\IEEEauthorrefmark{1},
Weiyi Liu\IEEEauthorrefmark{1},
Yi Li\IEEEauthorrefmark{1},
}
\IEEEauthorblockA{\IEEEauthorrefmark{1}Tsinghua Shenzhen International Graduate School, Tsinghua University, Shenzhen, 518055, China}
\IEEEauthorblockA{\IEEEauthorrefmark{2}College of Information Science and Electronic Engineering, Zhejiang University, Hangzhou, 310000, China}
Email: liyi@sz.tsinghua.edu.cn
}

\maketitle

\begin{abstract}

The Information Updating Networks (IUNs) offers significant potential for ocean exploration but encounters challenges due to dynamic underwater environments and severe system attenuation. Current methods relying on Autonomous Underwater Vehicles (AUVs) based on online reinforcement learning (RL) lead to high computational costs and low data utilization. To address these issues and the constraints of turbulent ocean environments, we propose a multi-AUV assisted data collection framework for IUNs based on multi-agent offline RL. This framework maximizes data rate and the value of information (VoI), minimizes energy consumption, and ensures collision avoidance by utilizing environmental and equipment status data. We introduce a semi-communication decentralized training with decentralized execution (SC-DTDE) paradigm and a multi-agent independent conservative Q-learning algorithm (MAICQL) to effectively tackle the problem. Extensive simulations demonstrate the high applicability, robustness, and data collection efficiency of the proposed framework.

\end{abstract}



\section{Introduction}\label{se:1}

The rapid development of marine resource exploration is constrained by the vastness of the ocean \cite{khalil2020toward}. Advances in information-routing communication have enabled the Information Updating Networks (IUNs), connecting underwater devices for applications like sensing, navigation, and environmental monitoring   \cite{xjzh1}. However, IUNs faces challenges such as severe system attenuation and the influence of ocean currents on data transmission \cite{senel2015self}. Moreover, information updates communication (IUC) is limited by high cost, narrow bandwidth, high bit error rates, slow speeds, and high energy consumption \cite{amar2016low}. Therefore, designing an efficient and reliable IUNs data collection scheme is essential.

Traditional data collection methods, such as opportunistic routing (OR) \cite{hsu2015delay}, K-means and ANOVA-based clustering \cite{harb2015enhanced}, and greedy routing protocols \cite{han2015routing}, face limitations like high communication overhead and failure risks. Autonomous Underwater Vehicles (AUVs) have proven effective in IUNs applications due to their mobility, endurance, and intelligence, balancing node energy consumption and extending network life \cite{kartha2017network}. However, most studies focus on single AUV methods \cite{zhu2017biologically}, which face path planning challenges.  Multi-AUV systems and cooperative control have been proposed to address these issues, but practical challenges remain, such as AUVs’ limited ability to communicate and cooperate \cite{lin2020path}.

Reinforcement learning (RL), particularly multi-agent RL (MARL), has been explored to enhance AUV data collection. Various studies have employed RL algorithms for AUV trajectory planning and data collection \cite{9978925, zhang2024environment}, but online RL faces challenges like high costs and potential AUV damage. Offline RL, using pre-collected datasets, improves learning efficiency and reduces interaction risks \cite{levine2020offline}. Multi-agent offline RL enhances cooperation and competition handling, distributed computation, learning efficiency, and exploration-exploitation balance without direct environment interaction, enabling better training outcomes for AUV clusters.

Based on above analysis, we propose a multi-AUV data collection framework using the multi-agent independent conservative Q-learning (MAICQL) algorithm within multi-agent offline RL. This framework optimizes objectives under environmental constraints, improving data utilization rates and IUNs task efficiency. Simulation experiments demonstrate that our framework significantly enhances data collection efficiency, robustness, and scalability across varying environments and numbers of AUVs.

The paper is structured as follows: Section \ref{se:2} presents the system model, including the scenario, model assumptions, and basic principles for the data collection task. Section \ref{se:3} formulates the data collection task as a multi-objective optimization problem with constraints. Section \ref{se:4} detail simulation experiments to evaluate the performance of the proposed framework, followed by conclusions presented in  Section \ref{se:5}.

\section{System Model}\label{se:2}
\subsection{Overview of the Data Collection Model}
\begin{figure}[t]
\centering
\includegraphics[width=0.948\linewidth]{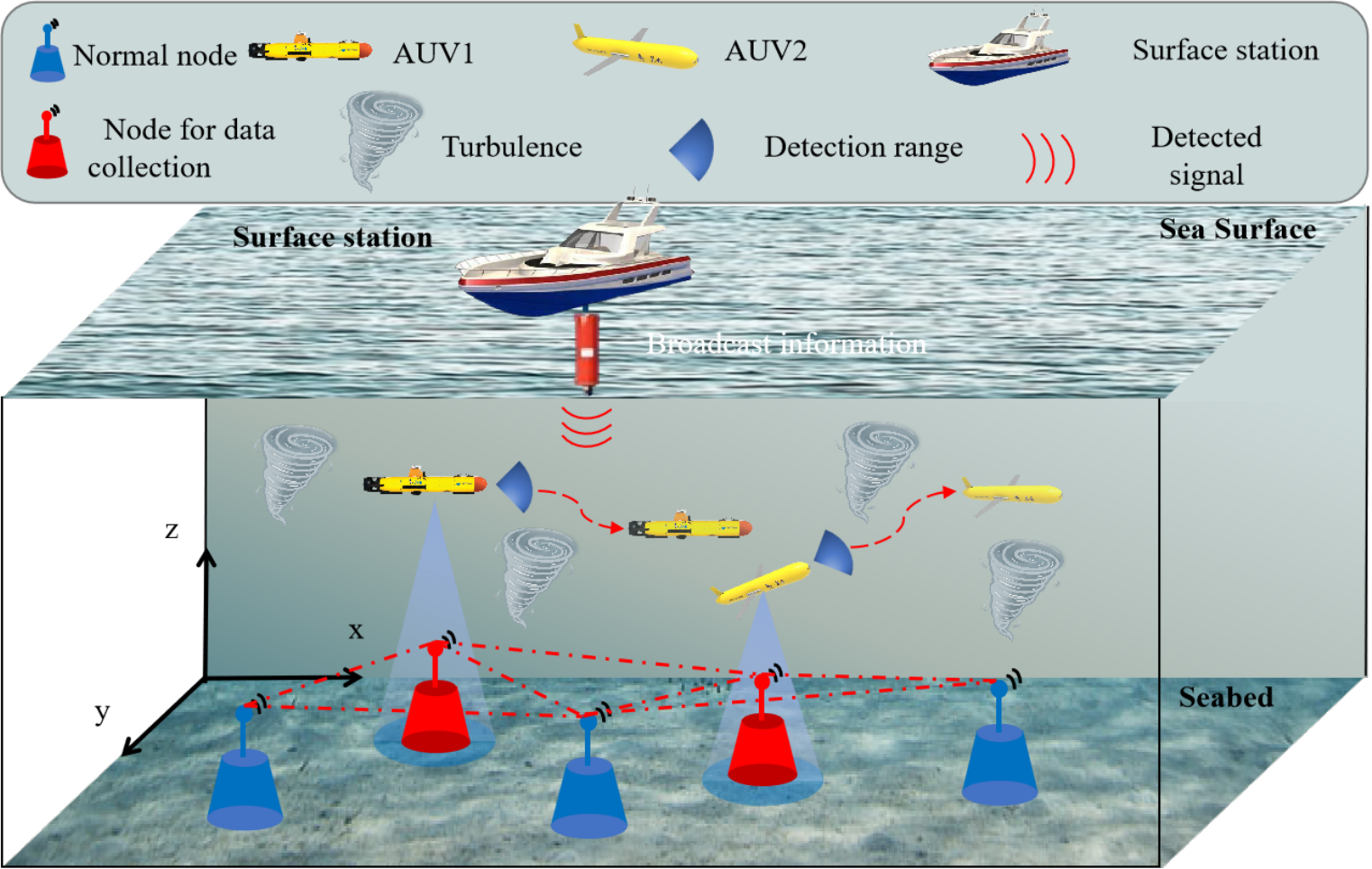}
\centering
\caption{Illustration of multi-AUV assisted IUNs data collection system.}
\label{fig:1}
\end{figure}

The multi-AUV assisted IUNs data collection framework developed in this study is illustrated in Fig. 1. The environment contains 
$N$ task-performing AUVs, denoted by $AUVs=\{AUV_1,AUV_2,\ldots,AUV_N\}$, and 
$\lambda$ IUNs nodes, represented by set $\Phi=\{\Phi_{1},\Phi_{2},\ldots,\Phi_{\lambda}\}$. These IUNs nodes include 
$\mu$ ordinary nodes $\Phi_{k}^{\mathrm{N}}=\{\Phi_{k_{1}}^{\mathrm{N}},\Phi_{k_{2}}^{\mathrm{N}},\ldots,\Phi_{k_{\mu}}^{\mathrm{N}}\}$, which do not require data collection, and $\nu$ data-collecting nodes $\Phi_{j}^{\mathcal{F}}=\{\Phi_{o_{1}}^{\mathcal{F}},\Phi_{o_{2}}^{\mathcal{F}},\ldots,\Phi_{o_{\nu}}^{\mathcal{F}}\}$, which do require data collection due to data backlog. AUVs receive status information of IUNs nodes from the base station on the sea surface to identify nodes needing data collection and work collaboratively to locate these nodes. They must avoid turbulent areas during operation to optimize their trajectories due to the significant impact of ocean turbulence on their movement and energy consumption.

It is essential to determine the priority based on the urgency of data collection required by each node. We define the channel capacity $\mathcal{N}_{\hbar}(t)\in[0,\mathcal{N}_{\hbar}^{max}]$ of a node $\hbar$ at time $t$ as the metric to assess the urgency of data collection for the node, where $\mathcal{N}_{\hbar}^{max}$ represents the maximum channel capacity of node $\hbar$. The priority $\mathcal{Q}_{\hbar}^{j}(t)$ of data collection from a node depends on the data accumulation ratio and the distance between node $\hbar$ and AUV $j$, which can be expressed as

\begin{equation}\mathcal{Q}_{\hbar}^{j}(t)=\frac{\mathcal{C}_{\hbar}(t)}{\mathcal{C}_{max}(\mathcal{N}_{\hbar}(t)+\varepsilon)}-\xi d_{\hbar}^{j}(t),\end{equation}
where $\mathcal{C}_\hbar(t)\in[0,\mathcal{C}_{max}]$ represents the data storage amount of node $\hbar$ at time $t$, $\mathcal{C}_{max}$ denotes the maximum data storage capacity of the node, $d_{\hbar}^{j}(t)$ represents the relative distance between AUV $j$ and node $\hbar$ at time $t$, $\xi$ is the distance penalty factor, and $\varepsilon$ denotes a constant to prevent calculation errors when $\mathcal{N}_{\hbar}(t)$ equals zero.

\subsection{AUV Data Channels Model}
In the data collection process, the AUV uses its onboard model for detecting ocean device nodes and for communication between AUVs. These processes are modeled using the underwater environment model equation
\begin{equation}EM=SL+TS+DI-NL-DT-2TL(d,f),\end{equation}
where $SL$, $TS$, $DI$, $NL$, and $TL(d,f)$ represent the output level, object strength, directional index, ambient noise level, and directional index, respectively. $DT$ and $EM$ denote the active model response threshold  and the echo margin, respectively. $TL(d,f)$ is related to the response radius  $d$ and the center parameter $f$, satisfying the following equation
\begin{equation}TL(d,f)=20\mathrm{log}(d)+\frac{d\kappa(f)}{1000},\end{equation}
where $\kappa(f)$ represents the absorption coefficient, calculated according to the Thorp formula
\begin{equation}\kappa(f)=0.11\frac{f^{2}}{1+f^{2}}+44 \frac{f^{2}}{4100+f^{2}}+2.75\times10^{-4}f^{2}+0.003.\end{equation}

In the underwater environment, the total noise $N_l$ consists of flow-field noise $N_t$, platform noise $N_s$, wind noise $N_w$, and thermal noise $N_{th}$, which can be represented by Gaussian statistics, and the total power spectral density (PSD) of $N_l$ is
\begin{equation}N_L(f)=N_t(f)+N_s(f)+N_w(f)+N_{th}(f).\end{equation}

The noise component in Eq. (5) can be expressed as
\begin{equation}
\left\{\begin{array}{l}\! \! \! 
10 \log N_t(f)\! =\! 17\! -\! 30 \log f, \\
\! \!\! 10 \log N_s(f)\! =\! 30\! +\! 20 s+\log \left(f^{26} /(f+0. 03)^{60}\right), \\
\! \!\!10 \log N_w(f)\! =\! 50\! +\! 7. 5 \omega^{1 / 2}\! +\! 20 \log \left(f /(f\! +\! 0. 4)^2\right), \\
\! \!\!10 \log N_{t h}(f)\! =\! -15+20 \log f,
\end{array}\right. 
\end{equation}
where $s\in(0,1)$ represents the activity factor, and $\omega$ is the wind speed with the unit of m/s.

\subsection{AUV Energy Consumption Model}
According to the empirical formula, the corresponding relationship between thrust $F_T$ and propulsion power $P$ is:
\begin{equation}\label{eq:9}
{F_T} =  - 0.0021{P^2} + 0.6342P + 2.8372.
\end{equation}

As the motor speed increases, the mechanical efficiency will gradually increase, the relationship between the sailing speed $v$ and the working efficiency $\eta$ of the propeller is:
\begin{equation}\label{eq:10}
\eta  = {F_T}v/P =  - 0.081{v^3} + 0.215{v^2} - 0.01v + 0.541.
\end{equation}

\subsection{Definition of the Value of Information}
The VoI is introduced to measure the relationship between the importance and timeliness of the collected data, which is defined as follows
\begin{equation}VoI_\hbar(t)=\begin{cases}\beta V_\hbar+(1-\beta)V_\hbar\mathcal{D}(t),&\quad t\geq T_\hbar,\\0,&\quad \rm otherwise,\end{cases}\end{equation}
where $\beta$ denotes the parameter that quantifies the significance and timeliness of collected data. $V_\hbar$ is the initial VoI for node $\hbar$, while $T_\hbar$ represents when the AUV begins data collection at this node. The function $\mathcal{D}(t)=e^{-(t-T_i)/\sigma}$ decreases over time, where $\sigma$ is a scaling factor. VoI diminishes as data transmission progresses, completing when data collection ends. $t_{c}^{\hbar}$ is the time the AUV spends collecting data at node $\hbar$, and $t_{m}^{\hbar}$ denotes the time taken to travel to next node. Then
\begin{equation}t_c^\hbar+t_m^\hbar=T_{\hbar+1}-T_\hbar.\end{equation}

Therefore, if AUV $j$ completes data collection at node $\hbar$ at time $t_\hbar$, moves to node $\hbar$+1, and starts data collection at node $\hbar$+1 at time $T_{\hbar+1}$, then the VoI of node $\hbar$ is updated as
\begin{equation}VoI_{\hbar}(T_{\hbar+1})=\beta E_{\hbar}+(1-\beta)E_{\hbar}e^{-\left(t_{c}^{\hbar}+t_{m}^{\hbar}\right)/\sigma},\end{equation}
where $t_c^\hbar=\frac{\mathcal{K}_\hbar(t)}{\mathcal{N}_\hbar(t)}$, while $t_\mathrm{m}^\hbar=\frac{l_{\hbar,\hbar+1}}{\overline{v_k}(\mathcal{r}_j^\hbar)}$, $\mathcal{K}_\hbar(t)$ represents the amount of data to be collected by node $\hbar$ at time $t$, $\mathcal{N}_\hbar(t)$ denotes the channel capacity of node $\hbar$ at time $t$, and $l_{\hbar,\hbar+1}$ signifies the distance between node $\hbar$ and node $\hbar$+1.

\subsection{Turbulent Ocean Environment Model}
In underwater environments, one of the primary challenges faced by AUVs is the impact of turbulent flow fields. The numerical equation of the ocean current model is represented by the superposition of several viscous vortex functions, which is described as follows

\begin{subequations}
\begin{align}
\!\!\!\!\!\!V_x(\boldsymbol{P}(t))\! \! =\! \! -\Gamma &\cdot \frac{y-y_0}{2 \pi\left\|\boldsymbol{P}(t)\! \! -\! \! \boldsymbol{P}_0\right\|_2^2} \cdot\left(1\! \! -\! \! e^{-\frac{\left\|\boldsymbol{P}(t)-\boldsymbol{P}_0\right\|_2^2}{\delta^2}}\right),\\
\!\!\!\!\!\!V_y(\boldsymbol{P}(t))\! \! =\! \! -\Gamma &\cdot \frac{x-x_0}{2 \pi\left\|\boldsymbol{P}(t)\! \! -\! \! \boldsymbol{P}_0\right\|_2^2} \cdot\left(1\! \! -\! \! e^{-\frac{\left\|\boldsymbol{P}(t)-\boldsymbol{P}_0\right\|_2^2}{\delta^2}}\right),\\
&\varpi(\boldsymbol{P}(t))=\frac{\Gamma}{\pi \delta^2} \cdot e^{-\frac{\left\|\boldsymbol{P}(t)-\boldsymbol{P}_0\right\|_2^2}{\delta^2}},
\end{align}
\end{subequations}
where $\boldsymbol{P}(t)$ and $\boldsymbol{P}_0$ are the current position of AUV and the coordinate vector of Lamb vortex center, $\delta$ is the radius of vortex, and $\Gamma$ is the intensity of vortex. Then the velocity of AUV $k$ under ocean current interference is
\begin{equation}\label{eq:28}
\boldsymbol{v}_k\left(\boldsymbol{P}_k({t})\right)=\boldsymbol{v}_k-\boldsymbol{v}_c\left(\boldsymbol{P}_k({t})\right),
\end{equation}
where $\boldsymbol{v}_c\left(\boldsymbol{P}_k({t})\right)$ is the water flow velocity at position $\boldsymbol{P}_k({t})$ and $\boldsymbol{v}_k\left(\boldsymbol{P}_k({t})\right)$ is the relative velocity at position $\boldsymbol{P}_k({t})$.

\section{Problem Formulation}\label{se:3}

\subsection{Optimization Problem Formulation}
There are multiple objectives and performance metrics to be optimized, which can be expressed as:

\begin{equation}\max_{\boldsymbol{v}^t}\mathcal{M}_r\!=\!\mathcal{L}\sum_{j=1}^NS_j^\mathcal{K}\!+\!\mathcal{J}\sum_{j=1}^NS_j^\mathcal{N}+\varsigma\sum_{j=1}^NS_j^\mathcal{V}-\mathcal{W}\sum_{j=1}^NE_j,\end{equation}
\begin{equation}\sum_{j=1}^N\!\chi_{j,i}\!=\!1, \forall i\!\in\! M,
 \!\sum_{i=1}^M\!\chi_{j,i}\!=\!1, \forall j\!\in\! N,
\boldsymbol{v}^t\!\in\![0,\boldsymbol{v}^t{}_{max}],\end{equation}
where $S_j^{\mathcal{K}}=\sum_{i=1}^M\sum_{\hbar=1}^{\Phi_0^{\mathcal{F}_j}}\chi_{j,i}[\hbar]\mathcal{K}_i(t_{\hbar,i})$ denotes the total data collection amount, $S_j^N=\sum_{i=1}^M\sum_{\hbar=1}^{\Phi_0^{\mathcal{F}_j}}\mathcal{N}_\hbar(t_{\hbar, i})$ represents the total channel capacity, and $\mathcal{N}_{\hbar}(t_{\hbar, i})\in[0,\mathcal{N}_{\hbar}^{max}]$ is the channal capacity of node $\hbar$ at time $t_{\hbar, i}$. $S_j^\mathcal{V}=\sum_{i=1}^M\sum_{\hbar=1}^{\Phi_0^{\mathcal{F}_j}}VoI_\hbar(t_{\hbar, i})$ is the total VoI for $j$-th AUV during an episode, and $t_{\hbar, i}$ represents the hover time of the AUV at the $i$-th node during the $\hbar$-th hover event. The factors $\mathcal{L}, \mathcal{J}, \varsigma, \mathcal{W} \in (0,1)$ denote the contribution factors for weighting these objectives. The constraints ensure only one AUV hovers at a node at a time, and each AUV hovers over only one node at a time. The AUV's motion state must stay within the range defined by $\boldsymbol{v}_{max}^{t}=[v_{x_{max}}^{t},v_{y_{max}}^{t},\varpi_{max}^{t}]^{T}$.

\subsection{Markov Decision Process Model}
We model the data collection process as an MDP, which can be represented by a quintuple
\begin{equation}\begin{split}
\epsilon=\{[\boldsymbol{\mathcal{S}}_1^t,\boldsymbol{\mathcal{S}}_2^t,\ldots,\boldsymbol{\mathcal{S}}_N^t],[\boldsymbol{A}_1^t,\boldsymbol{A}_2^t,\ldots,\boldsymbol{A}_N^t],\boldsymbol{P},\\ 
[R_1(t),R_2(t),\ldots,R_N(t)],\gamma\},
\end{split}\end{equation}
where $\boldsymbol{\mathcal{S}}=[\boldsymbol{\mathcal{S}}_1^t,\boldsymbol{\mathcal{S}}_2^t,\ldots,\boldsymbol{\mathcal{S}}_N^t]$, $\boldsymbol{A}=[\boldsymbol{A}_1^t,\boldsymbol{A}_2^t,\ldots,\boldsymbol{A}_N^t]$, $\boldsymbol{P}$, $\boldsymbol{R}=[R_1(t),R_2(t),\ldots,R_N(t)]$ respectively represent state space, action space, state transition probability distribution and reward function, and $\gamma$ is the discount factor. Each parameter is specified as follows: 

\textbf{State Space $\boldsymbol{\mathcal{S}}$}: The information observed by each AUV includes its own state, detectable partial environmental information, and the state information of other AUVs. Therefore, the state space of AUV $j$ at time $t$ can be defined as
\begin{equation}\boldsymbol{\mathcal{S}}_j^t=\begin{Bmatrix}\mathcal{I}_j(t),\mathcal{I}_j^{N-1}(t),\mathcal{E}_j^M,\mathcal{G}_j^t(\boldsymbol{r})\end{Bmatrix},\end{equation}
where $\mathcal{I}_j(t)$ and $\mathcal{I}_j^{N-1}(t)$ represent the state information of AUV $j$ and other AUVs at time $t$ respectively, $\mathcal{E}_j^M$ is the environmental information detectable by AUV $j$, and $\mathcal{G}_j^t(\boldsymbol{r})$ represents the position and velocity of the turbulence relative to AUV $j$ at the current time.

\textbf{Action Space $\boldsymbol{A}$}: The actions that AUVs can take include object node selection, acceleration, and angular velocity. Thus, the action space of AUVs can be represented as
\begin{equation}\boldsymbol{A}_j^t=\begin{Bmatrix}\mathcal{P}_j(t),a_j(t),\varpi_j(t)\end{Bmatrix},\end{equation}
where $\mathcal{P}_j(t)$ is the set of nodes selected by the AUV based on the priority, $a_j(t)$ represents the acceleration, and $\varpi_j(t)$ denotes the angular velocity.

\textbf{Reward Function $R$}: The agent assesses its action policies based on the rewards from the environment. Thus, designing an appropriate reward function is crucial for enhancing data collection efficiency. The reward function can be expressed as:
\begin{equation}
    R_j(t)\!=\! w^{\text{ec}}r^{\text{ec}}_j(t)\!+\!w^{\text{VoI}}r^{\text{VoI}}_j(t)\!+\!w^{\text{dr}}r^{\text{dr}}_j(t)\!+\!w^{\text{dp}}r^{\text{dp}}_j(t)\!+\!w^{\text{cs}}r^{\text{cs}}_j(t),
\end{equation}
where $r^{\text{ec}}_j(t)$ denotes the energy consumption of $j$-th AUV of time $t$, $r^{\text{VoI}}_j(t) = \sum_{i=1}^M\sum_{\hbar=1}^{\Phi_o^{\mathcal{F}_j}}VoI_\hbar(t_{\hbar, i})$ denotes the total obtained VoI, $r^{\text{dr}}_j(t)=\sum_{i=1}^M\sum_{\hbar=1}^{\Phi_0^{Fj}}\chi_{j,i}[\hbar] \mathcal{K}_i(t_{\hbar,i})$ represents the total data rate, and $r^{\text{cs}}_j(t)=\sum_{i=1,i\neq j}^N(1-\max{(d_{j,i},d_s)/d_s})/(N-1 )$ represents the negative reward to penalize collisions. Besides, he reward term $r^{\text{dp}}$ encouraging priority data collection is specified by
\begin{equation}r^{\text{dp}}_j(t)=\begin{cases}\frac{1}{(d_j^i(t)+o)},&d_j^i(t)\leq d_r,\\0,&\rm otherwise,\end{cases}\end{equation}
where $d_j^i(t)$ is the distance between the AUV $j$ and the object node $i$, $d_r$ is the distance threshold, and $o$ is a constant to prevent calculation error when $d_j^i(t)=0$.

\section{Algorithm Design}\label{se:4}

\subsection{Semi-communication DTDE Model}

Centralized training with decentralized execution (CTDE) relies on communication between agents and a central controller, leading to delays that hinder training and decision-making. In contrast, fully DTDE treats agents independently, improving training efficiency but often resulting in suboptimal performance. To address these issues, we propose the SC-DTDE model, where each AUV has partial knowledge of other AUVs' states without requiring complete environmental information. Denoting the observed state of AUV $j$ at time $t$ as $o_j^t$, the obtained state information can be expressed as
\begin{equation}s_j^t=\{e_j^1o_1^t,e_j^2o_2^t,\ldots,e_j^No_N^t\},\end{equation}
where $e_j^N$ denotes the weight parameter associated with the status information of AUV $N$, as perceived by AUV $j$. In SC-DTDE, the AUV policies operate independently, without shared parameters. Upon the completion of the training phase, the AUV dispenses with the value network $Q(o_j^t;\mathcal{s}_j)$, relying solely on the policy network $\pi(a_j^t\mid o_j^t;\theta_j)$, which is integrated into its own system, for decision-making processes. This decision-making framework is partially communicative. The employed SC-DTDE model is not only expedient but also fosters inter-AUV connectivity and collaboration.

\subsection{Offline Dataset Collection}

Offline RL uses pre-existing datasets from expert performances, avoiding direct interaction during training.  This is beneficial for training AUVs in IUNs data collection. 

For dataset generation, we first employ soft actor-critic (SAC), the classical online RL algorithm for policy improvement of AUVs. Then the policy with expert-level is selected to be deployed in the AUVs in the turbulent ocean environment to generate the offline dataset, which includes 400 epochs' data. Utilizing the SC-DTDE framework and the dataset, the MAICQL algorithm is then employed to train and optimize the policies of multi-AUV, which will be detailed in the subsequent subsection.

\begin{algorithm}[!t]
\label{alg:2}
\caption{MAICQL Algorithm}
Initialize the replay buffer $\mathcal{D}$, entropy regularity coefficient $\alpha$, critic network, target critic network, and policy network parameters $\varepsilon_{1j}$, $\varepsilon_{2j}$, $\varepsilon_{1j}^{-}$, $\varepsilon_{2j}^{-}$, $\theta_j$ of AUV $j$.

Import the expert dataset $\mathbb{T}$.

\For{each epoch $k$}{
Reset the training environment and parameters.

\For{each time step $t$}{

\For{each AUV $j$}{
Extract $N$ tuples of data ${(\boldsymbol{s}_n,\!{a}_{1_n}\!,\!\cdots\!,{a}_{N_n},R_{t_n},\boldsymbol{s}_n^{\prime})}_{n\!=\! 1,\!\cdots\!,N}$ \!from $\mathbb{T}$. 

Update the entropy regularity coefficient:
\begin{equation*}\begin{aligned}
&\alpha_t\leftarrow\alpha_{t-1}-\\&\eta_\alpha\!\nabla_{\!\alpha}\mathbb{E}_{s\sim \mathcal{D},a\sim\pi_\theta^j(a|s)}[-\alpha_{t\!-\! 1}\mathrm{log}\pi_\theta^j(a|s)\!-\!\alpha_{t\!-\! 1}\mathcal{H}].\end{aligned}\end{equation*}

Update the critic networks $Q_{1j}$ and $Q_{2j}$ according to Eq. (24).

Update the target critic networks:
\begin{equation*}\varepsilon_{1t}^{-}=\tau\varepsilon_1^t+(1-\tau)\varepsilon_{1}^{t-1},\varepsilon_{2t}^{-}=\tau\varepsilon_2^t+(1-\tau)\varepsilon_{2}^{t-1}.\end{equation*}

Update the policy network:
\begin{equation*}\begin{aligned}&\theta_{t}\!\!\leftarrow\!\!\theta_{t\!-\! 1}\!\!-\!\!\eta_{\pi}\!\!\nabla_{\!\theta}\mathbb{E}_{s\sim\mathcal{D},a\sim\pi_{\theta-1}^{j}\!(a|s)}[\alpha\mathrm{log}\pi_{\theta\!-\! 1}^{j}\!(a|s)-\\&\mathrm{min}(Q_{1j}^{\varepsilon_{1}}(s,a),Q_{2j}^{\varepsilon_{2}}(s,a))].\end{aligned}\end{equation*}
}
}}
\end{algorithm}

\subsection{Multi-Agent Independent Conservative Q-Learning Algorithm}

Offline RL must address the challenge of minimizing extrapolation error, which often leads to overestimation of the value function at points far from the dataset. In this work, we extend the CQL algorithm to its multi-agent variant, denoted as MAICQL. We apply MAICQL to train multiple AUVs simultaneously and independently through SC-DTDE. Addition to the temporal difference (TD) target of online RL methods, the policy-derived actions are also penalized according to their alignment with state data from the expert dataset, thereby refining the learning process
\begin{equation}\begin{aligned}\widehat{Q_{j}^{\varepsilon+1}}&\leftarrow \mathrm{argmin}_Q\alpha\mathbb{E}_{s\sim\mathbb{D},a\sim\pi_{\theta}^{j}(a|s)}[Q(s,a)]\\&+\frac{1}{2}\mathbb{E}_{(s,a)\sim\mathbb{D}}\left[\left(Q(s,a)-\widehat{\mathbb{B}}^{\pi}\widehat{Q_{j}^{\varepsilon}}(s,a)\right)^{2}\right],\end{aligned}\end{equation}
where $\widehat{\mathbb{B}}^{\pi}(s_{k},a_{k},s_{k+1},a_{k+1})$ is the Bellman operator of policy $\pi$, $\alpha$ is the entropy regularity coefficient, , $\mathbb{E}$ is the expectation, and $\mathbb{D}$ is the state-action space. However, it is not necessary to constrain all points within the dataset, as this approach may result in an unnecessary computational overhead. It is reasonable to assume that data points generated by the policy function that align with expert policy $\mu_\omega$ are inherently more accurate. Consequently, these points may not require constraints
\begin{equation}\begin{aligned}&\widehat{Q_{j}^{\varepsilon+1}}\leftarrow\\&\mathrm{argmin}\alpha\left(\mathbb{E}_{s\sim\mathbb{D},a\sim\pi_{\theta}^{j}(a|s)}[Q(s,a)]-\mathbb{E}_{s\sim\mathbb{D},a\sim\mu_{\omega}(a|s)}[Q(s,a)]\right)\\&+\frac{1}{2}\mathbb{E}_{(s,a)\sim\mathbb{D}}\left[\left(Q(s,a)-\widehat{\mathbb{B}}^{\pi}\widehat{Q_{j}^{\varepsilon}}(s,a)\right)^{2}\right].\end{aligned}\end{equation}

We further substitute $\pi^{j}_{\theta}$ with a policy function $\widehat{\pi}^{j}_{\theta}$ that maximizes the value function $Q$. Additionally, we introduce a regularization term $\mathbb{R}$ to mitigate overfitting. This term is quantified by the Kullback-Leibler divergence between the policy $\widehat{\pi}^{j}_{\theta}$ and a prior policy $\mathbb{U}$, which adheres to uniform distribution. Therefore, the complete target of the critic network is

\begin{figure*}[!t]
	
	\begin{minipage}{0.335\linewidth}
		\vspace{3pt}
		\centerline{\includegraphics[width=\textwidth]{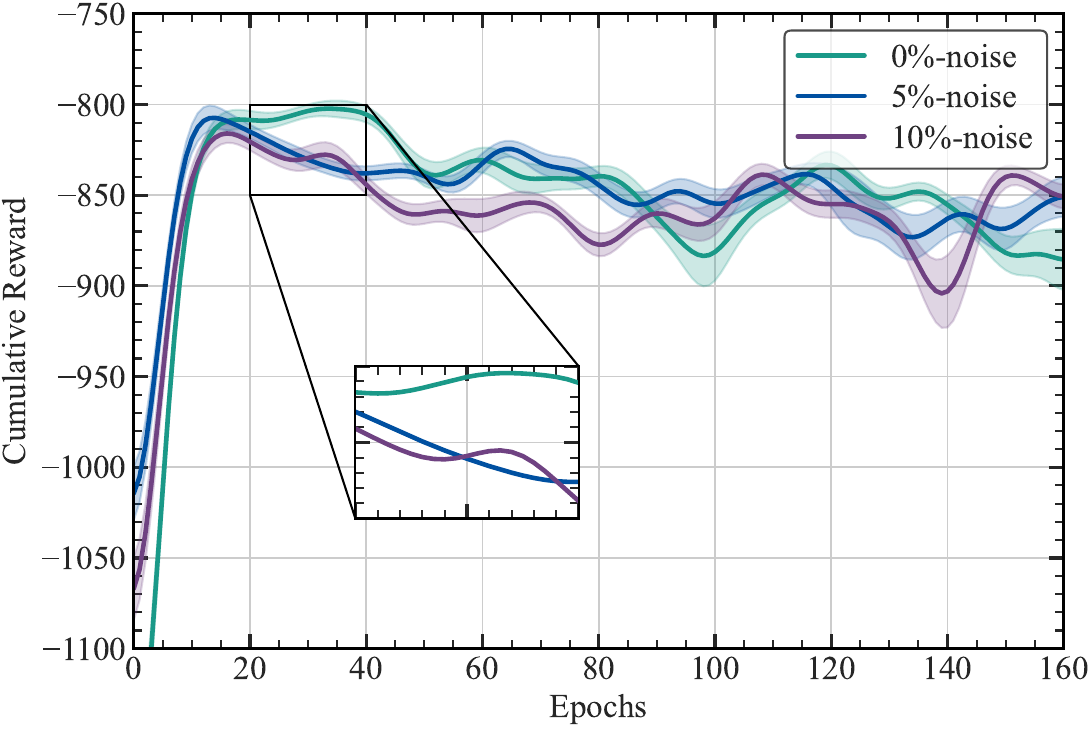}}
		\centerline{(a) Cumulative reward}
	\end{minipage}
	\begin{minipage}{0.32\linewidth}
		\vspace{3pt}
		\centerline{\includegraphics[width=\textwidth]{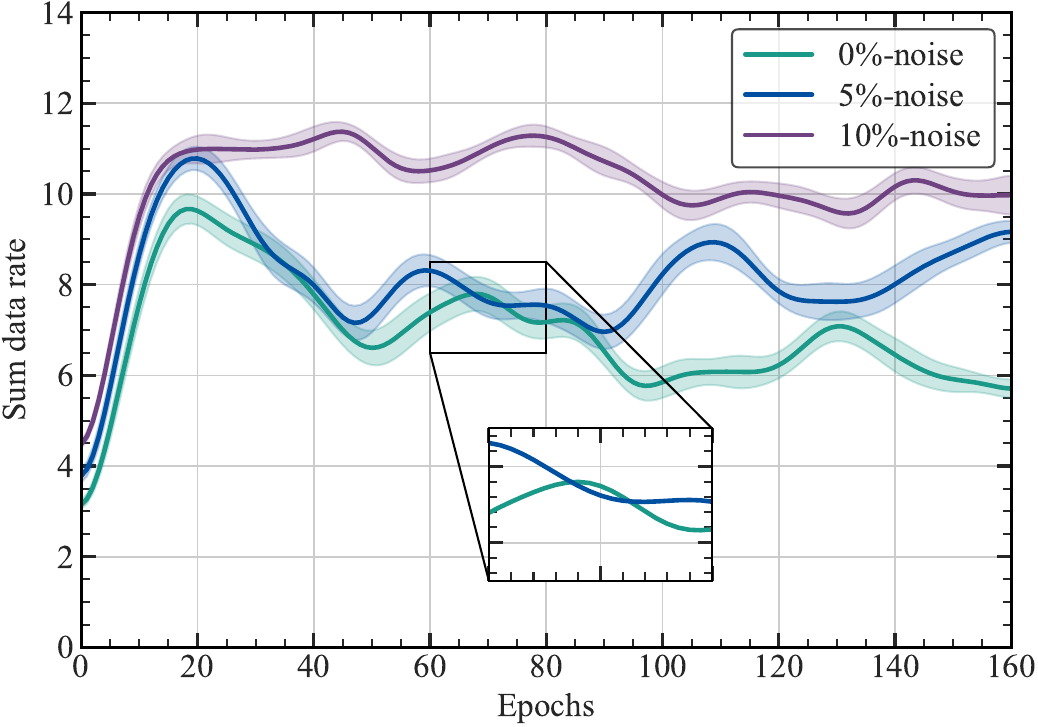}}
	 
		\centerline{(b) Sum info rate}
	\end{minipage}
	\begin{minipage}{0.325\linewidth}
		\vspace{3pt}
		\centerline{\includegraphics[width=\textwidth]{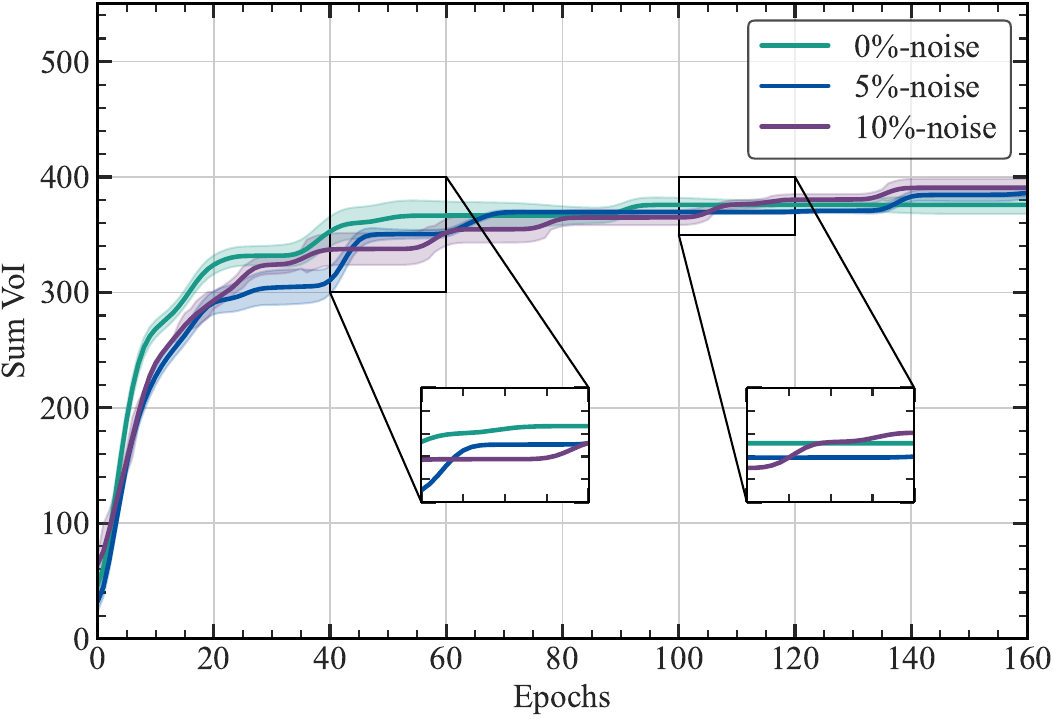}}
	 
		\centerline{(c) Sum VoI}
	\end{minipage}
 
	\caption{The curves of the cumulative reward, sum info rate, and sum VoI under different noise intensities: (a) Cumulative reward. (b) Sum info rate. (c) Sum VoI.}
	\label{fig:2}
\end{figure*}

\begin{equation}\begin{aligned}&\widehat{Q_{j}^{\varepsilon+1}}\leftarrow \\&\mathrm{argmin}_Q\alpha\mathbb{E}_{s\sim\mathbb{D}}\left[\mathrm{log}\sum_\alpha \exp\left(Q(s,a)\right)\!-\!\mathbb{E}_{a\sim\mu_\omega(a|s)}[Q(s,a)]\right]\\&+\frac12\mathbb{E}_{(s,a)\sim\mathbb{D}}\left[\left(Q(s,a)-\widehat{\mathbb{B}}^\pi\widehat{Q_{j}^{\varepsilon}}(s,a)\right)^2\right].\end{aligned}\end{equation}

Additionally, in MAICQL, each AUV is equipped with two value functions $Q_{1j}$ and $Q_{2j}$, and a policy function $\pi^{\theta}_j$. Furthermore, we introduce dual objective functions, $Q^T_{1j}$ and $Q^T_{2j}$, designed to minimize overestimation during the updating. The full objectives of MAICQL are shown in Algorithm 1.

\section{Simulation Results and Discussions}\label{se:5}

\subsection{Experiment Settings}
The experimental code in this study was executed on a personal computer equipped with 13th Gen Intel® Core™ i7-13650HX processor and NVIDIA GeForce RTX 4060 Laptop GPU. Algorithm parameters, motion parameters, system communication parameters, and other parameters of our experiment are detailed in Table. \ref{tab:1}.

\begin{table}[!t]
\caption{Parameters of The Environment and Algorithm.\label{tab:1}}
\centering
\setlength{\tabcolsep}{0.5mm}{
\begin{tabular}{m{4.7cm}<{\centering}c}
\hline
{\bf Parameters} & {\bf Values}\\
\hline
Field size $S$ & $120\mathrm{m} \times 120\mathrm{m}$ \\
The number of IUNs devices $N_d$ & 55\\
Transmit parameter $f$ & 20$\mathrm{arb}$\\
Field depth $H$ & $-50 \mathrm{m}$\\
The number of AUVs $N$ & $2$\\
Motion depth  $d$ & -10 $\mathrm{m}$\\
Update range parameter $L$ & 6.0 $\mathrm{m}$\\
Maximum speed $v_{max}$ & 2$\mathrm{m/s}$\\
Intensity of turbulent vortex $\Gamma$ & 8\\
Turbulence radius $r$ & 48$\mathrm{m}$ \\
Training duration $T$ & 1000$\mathrm{s}$\\
Step size $\Delta T$ & 1.0$\mathrm{s}$\\
The neural network's learning rate $lr$ & 0.0002\\
The entropy regularity coefficient's learning rate $lr_\alpha$ & 0.0003\\
The entropy regularity coefficient's initial value $\alpha$ & 0.01\\
The soft update coefficient $\tau$ & 0.01\\
The discount factor $\gamma$ & 0.99\\
Training epochs & 400\\
The Bellman operator $\widehat{\mathbb{B}}^{\pi_{\theta}}$ & 5\\
Target entropy $\mathcal{H}_0$ & 2.0\\
The VoI scaling parameter $\sigma$ & 10\\
The VoI measure parameter $\beta$ & 0.7\\
Maximum data storage amount $\mathcal{C}_{max}$ & 2$\mathrm{Mbits}$\\
Crash distance $d_s$ & 5$\mathrm{m}$\\
Replay buffer size $\mathcal{D}$ & $8\times 10^4$\\     
\hline
\end{tabular}}
\end{table}

\subsection{Simulation Results and Analysis}
At the start of each epoch, the positions of AUVs and the device states are reset. Each AUV selects a object device using Eq. (1), navigates to collect data, and repeats this until the task is complete. The training effect is evaluated using the system's cumulative reward, sum info rate, and sum VoI.

By incorporating varying intensities of Gaussian noise into the dataset within the MAICQL algorithm, real underwater environments are more accurately simulated. The algorithm's performance under noisy conditions is analyzed, confirming its robustness. As shown in Fig. \ref{fig:2}, the system's three metrics improve during training, regardless of noise presence. Since introducing noise enhances performance by increasing state-action pairs near distribution edges, thus improving sampling efficiency, which validates MAICQL's robustness.

\begin{figure}[!t]
\centering
\includegraphics[width=0.888\linewidth]{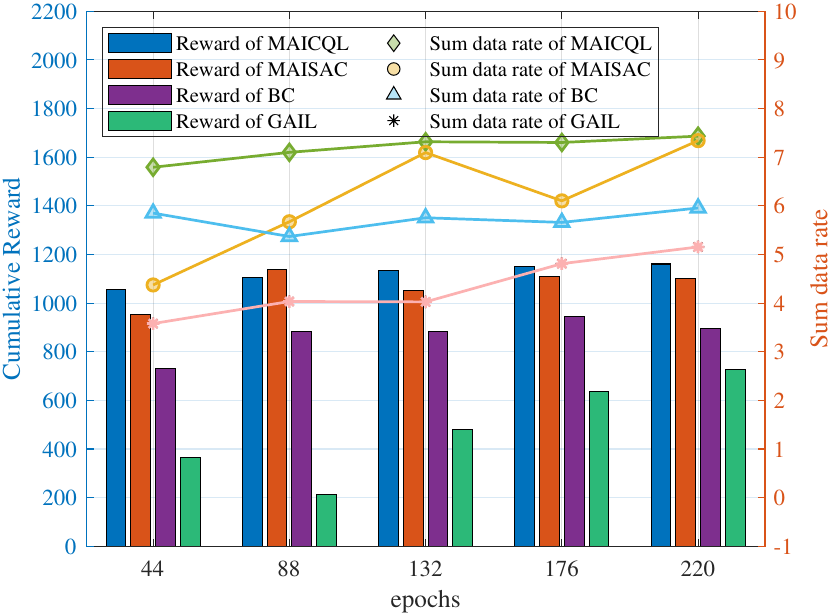}
\caption{Performance comparison of MAISAC, BC, GAIL and MAICQL algorithms.}
\label{fig:3}
\end{figure}

To evaluate the algorithm's performance in different environment settings, Table \ref{tab:2} shows system metrics changes with varying numbers of AUVs. More AUVs increase the sum info rate and reduce average energy consumption when increasing from 2 to 3 AUVs. However, this also lead to a higher average collision rate, raising the risk of damage. After weighing various factors, the optimal number of AUVs is determined to be 2.

\begin{table}[!t]
\centering
\caption{Comparison of various system metrics under different AUV number.\label{tab:2}}
\label{table:masking_performance}
\vspace{3mm}
\begin{tabular}{lccc}
\hline
Setting & \textbf{Sum info rate} & \textbf{Avg energy cost} & \textbf{Crash Number} \\ 
\hline
\textbf{N=1}    & 1.98 & 136.27 & 0 \\ 
\textbf{N=2}    & 11.51 & 160.01 & 6.03 \\ 
\textbf{N=3}    & 17.55 & 134.35 & 12.03 \\
\hline
\end{tabular}
\vspace{-1mm}
\end{table}

\begin{figure}[!t]
\centering
\includegraphics[width=0.788\linewidth]{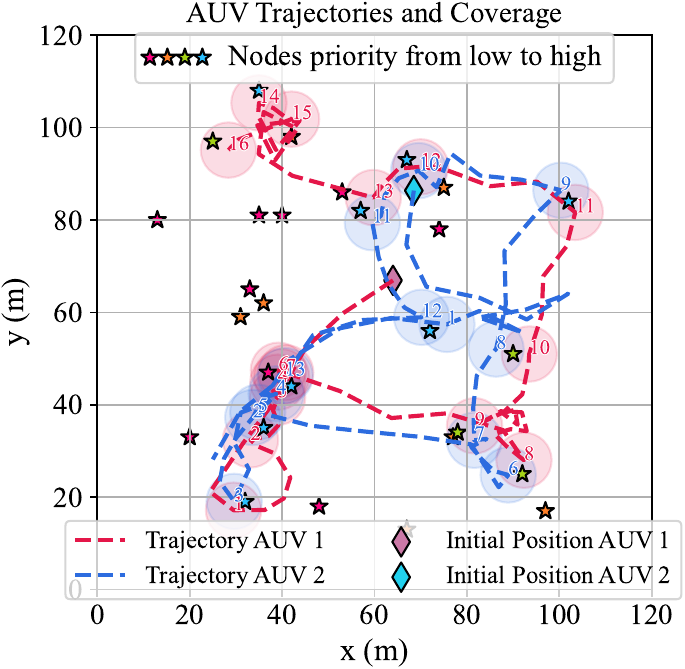}
\caption{Trajectories of AUVs for the data collection task in the turbulence-free environment.}
\label{fig:4}
\end{figure}

\begin{figure}[!t]
\centering
\includegraphics[width=0.848\linewidth]{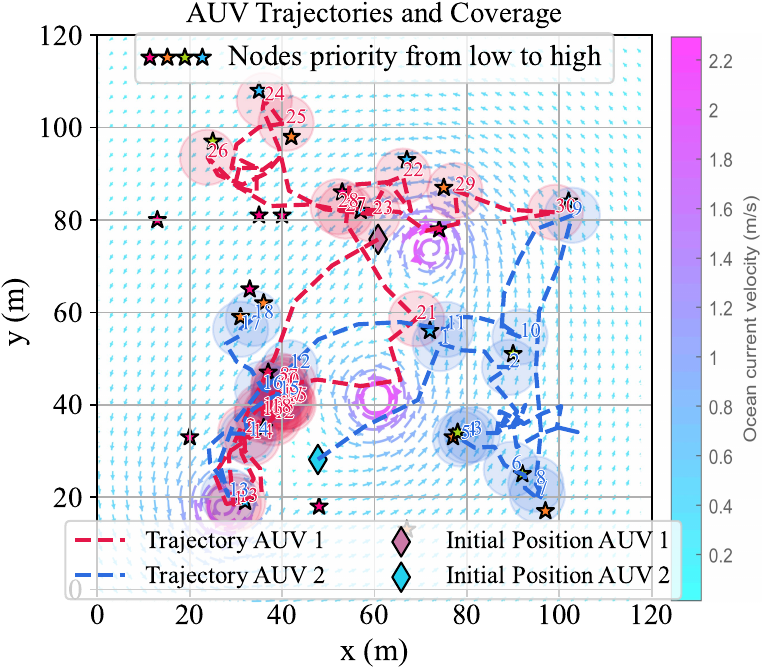}
\caption{Trajectories of AUVs for the data collection task in the turbulent environment.}
\label{fig:5}
\end{figure}

To demonstrate the advantages of the SC-DTDE-based MAICQL algorithm, we compare it with MAISAC \cite{zhang2024environment}, behavioral cloning (BC), and generative adversarial imitation learning (GAIL). As shown in Fig. \ref{fig:3}, with rewards shifted by 1200 for clarity, MAICQL consistently outperforms BC and GAIL in cumulative reward and slightly surpasses MAISAC. It also achieves a higher sum info rate and greater stability, highlighting its superior performance.

Finally, to validate the proposed IUNs data collection framework, two AUVs trained with MAICQL were tested in both turbulence-free and turbulent environments, and the trajectories of AUVs are shown in Fig. \ref{fig:4} and Fig. \ref{fig:5}. The AUVs consistently prioritized urgent, nearby nodes, enhancing efficiency and minimizing energy use. In turbulent environments, they effectively hovered and collected data at relevant nodes, demonstrating the framework's capability for robust multi-objective optimization in IUNs data collection.

\section{Conclusion}\label{se:6}

This paper presents an offline RL-based multi-AUV framework for IUNs data collection in turbulent ocean environments. The problem is modeled as an MDP with a strategically designed reward function to enhance decision-making. We then propose MAICQL algorithm, based on the SC-DTDE model, to improve data collection efficiency while minimizing environmental interaction. Extensive experiments demonstrate the framework's robustness and superior performance. Future work will involve sim2real experiments in real environments.

\bibliographystyle{IEEEtran}

\end{document}